# A guide to Value of Information methods for prioritising research in health impact modelling

R. Johnson, J. Woodcock, A. de Nazelle, T. de Sa, R. Goel, M. Tainio, and C. Jackson

Health impact simulation models are used to predict how a proposed intervention or scenario will affect public health outcomes, based on available data and knowledge of the process. The outputs of these models are uncertain due to uncertainty in the structure and inputs to the model. In order to assess the extent of uncertainty in the outcome we must quantify all potentially relevant uncertainties. Then to reduce uncertainty we should obtain and analyse new data, but it may be unclear which parts of the model would benefit from such extra research.

This paper presents methods for uncertainty quantification and research prioritisation in health impact models based on Value of Information (VoI) analysis. Specifically, we
1. discuss statistical methods for quantifying uncertainty in this type of model, given the typical kinds of data that are available, which are often weaker than the ideal data that are desired;
2. show how the expected value of partial perfect information (EVPPI) can be calculated to compare how uncertainty in each model parameter influences uncertainty in the output;
3. show how research time can be prioritised efficiently, in the light of which components contribute most to outcome uncertainty.

The same methods can be used whether the purpose of the model is to estimate quantities of interest to a policy maker, or to explicitly decide between policies. We demonstrate how these methods might be used in a model of the impact of air pollution on health outcomes.

## Introduction

To predict the impact on public health of an intervention or policy, simulation models are often used. Under the intervention or policy, one or more health outcomes (e.g. lung cancer, stroke) is modified, typically through modifying one or more risk factors (e.g. physical inactivity, air pollution). Models bring together multiple sources of data and assumptions, describe the mechanisms by which risk factors and outcomes are affected, and calculate one or more estimates of the overall health impact. A simplified example is given in Figure 1, drawn from a model (the "Integrated Transport and Health Impact Model (ITHIM)" (1)) that examines feasible changes in transport behaviours and/or policies. These changes are assumed to affect health through three pathways: physical activity related to active transport (e.g. bicycling and walking), exposure to air pollution, and road-traffic injuries.

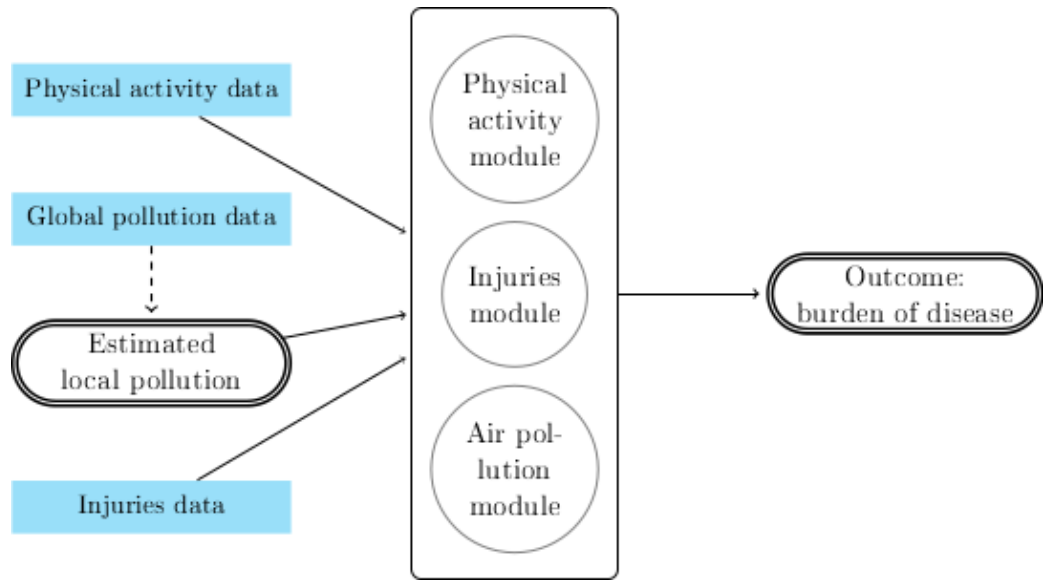

*Figure 1. Integrated Transport and Health Impact Model (ITHIM). The data are represented by blue nodes. The model defines the outcome (burden of disease) as a function of input data.*

Predictions from the models naturally have uncertainty. The processes being modelled can never be known perfectly, and the data informing the model are typically weaker than the "ideal" data desired, in terms of sample size, potential biases or relevance. However, it is possible to learn which inputs to the model most influence uncertainty in the outcome, and therefore prioritise further research or data collection.

Literature on quantifying uncertainty in decision models for health economic evaluation (e.g. (2-4)) recommends that, where possible, uncertainties are represented by model parameters, which are given probability distributions. We follow this framework, where a model is represented as a function $f$ mapping a set of uncertain quantities $X$ to a set of outputs $Y = f(X)$. A probability distribution is placed on $X$ to represent uncertainty about the inputs. The implied distribution on $Y$ is typically computed by Monte Carlo simulation. Constructing such functions and distributions can be particularly challenging in simulation models for public health impacts, since the interventions or scenarios of interest can affect multiple diseases, through multiple risk factors and pathways, thus requiring many sources of data and potential uncertainty to be parameterised (as reviewed in e.g. (5)).

This probabilistic framework forms the basis for Value of Information analysis, which has been used extensively in health economic decision modelling (6) to estimate the potential health economic benefits from gaining more information about an uncertain model quantity. However, these methods are underused in health impact modelling, where models can be used for broader purposes (7). In health economic decision models, the decision maker faces a choice between a discrete set of interventions and chooses to adopt the intervention with the maximum expected net benefit (8). In health impact modelling, the "output" is not necessarily a discrete *decision*, but is commonly a set of *estimates* of expected health outcomes under different potential circumstances. The theory of how VoI can be formulated and calculated for a model used for estimation rather than discrete decision making was described in (9), drawing ideas from sensitivity analysis in computer models (10,11), Bayesian experimental design (12), and VoI computation for discrete decisions (13).

In this paper, we show how VoI methods can be applied in health impact models. As using VoI methods depend on quantifying uncertainty probabilistically, we discuss how this can be done in typical situations. Most commonly, the available data are potentially biased or represent a slightly different setting compared to the ideal data. In the next section, we discuss a range of statistical models that can represent the relation of the observed data to the quantity of interest, while naturally producing a probabilistic representation of uncertainty. For example, hierarchical models can be used to inform parameters for which there are data from multiple related settings. In the following section, we describe VoI methods, and recommend a general procedure for how they might be used to prioritise further research to strengthen the evidence base of a typical health impact model.

Finally, we illustrate these ideas in an artificial example adapted from the air pollution module of ITHIM São Paulo (1, 14). The model calculates changes in expected deaths and disability under different scenarios of travel patterns. In this model, disease risk is modelled as a function of air pollution exposure; air pollution exposure is a function of background air pollution and ventilation rate; ventilation rate is a function of the individual's travel pattern, and background air pollution is a function of the population's travel patterns. Full details are given in the Appendix. We adapt the example from the version published in (14) to emulate a hypothetical new setting where an appropriate model structure has been agreed but some parameters are uncertain. We illustrate how probabilistic modelling of uncertainty, combined with Value of Information analysis, can help to identify the parts of the model where uncertainty has the greatest impact on the model output, thus to prioritise the available research time to better inform the model.

## Uncertainty quantification in health impact models

The uncertainty in a model output depends on how uncertainty about the model parameters is specified. In this section, we give general principles for quantifying common types of uncertainty in health impact models. The types of uncertainties we consider arise from incomplete knowledge of the processes being modelled or populations represented ('epistemic' uncertainties), which may in principle be reduced by obtaining more data. We quantify uncertainty as probability distributions for model parameters, to enable VoI methods to be used to prioritise further research. Statistical models are used to describe the relation between the observed data and the model parameter, and if Bayesian inference is used then a probability distribution representing plausible values for the parameter arises naturally.

We focus on a common challenge in health impact models: that the relevance and the quality of the observed data may vary. We briefly review potentially useful methods from the literature, and in the Appendix two of these methods are illustrated in the context of our example pollution model.  Note that in any of these methods, there may still be "structural" uncertainty about the choice of statistical models or distributions to employ – to handle these, in general, we recommend flexible models that ensure that all potential sources of uncertainty are characterised as parameters with probability distributions, as discussed in e.g. (4).

**Using published data directly**

The "gold standard" for informing a model parameter would be to use data that are representative and unbiased. This could be a published point estimate, with a confidence / credible interval or standard error that reflects the size of the population used to calculate the estimate. In these cases, statistical principles (see e.g. (3,6)) can provide an appropriate distribution around the parameter, for example one with standard deviation equal to the reported standard error. In principle, if we were to observe a larger population from the same source, the uncertainty would decrease.

**Using reported estimates of parameters directly with inflated variance**

If published parameter estimates are nominally representative, i.e. measure directly the quantity of interest, but are potentially biased due to a small sample, flawed study conduct, or insufficient documentation (including data that report only point estimates without uncertainties), then we might choose to use a distribution with a greater variance than the reported one.  This is equivalent to replacing a model parameter X by X+d, where the distribution for X comes from the published data, and d is a bias term whose mean and variance depend on the expected amount and direction of bias, which would require a case-by-case judgement (see e.g. 27,28).

**Using published data indirectly**

In the absence of data on our parameter of interest, we can use data that nominally represent something slightly different from what we require, for example data from a similar but not identical population, time period or place, or measurements of a slightly different but related outcome.  It is a matter of judgement whether the data are "similar enough" to be considered direct estimates of the parameter, if they represent the quantity of interest but with an unknown degree of bias, or if their difference from what is required can be explained.  If the difference can be explained, instead of arbitrarily inflating a variance, we can specify a statistical model that defines the relationship between the data and the parameter of interest. Bayesian inference naturally produces a quantification of uncertainty as the posterior distribution for the parameter.

Health impact models often involve risk factors, such as diet, physical activity and pollution, which are hard to define and measure.  How to quantify uncertainty depends on the purpose of the analysis. For air pollution, for example, the available data may be at the wrong level of aggregation, as well as for the wrong place and time: an area-level average may only be available where an individual-level local measurement is wanted ("Berkson" measurement error), or vice versa ("classical" error), or there may be uncertainty in how to combine observed data with model outputs (15). In each case, developing a statistical model for how all quantities are related will lead to a better reflection of uncertainty (see, e.g (15-18)). For physical activity, reported units of measurement may vary, and synthesising all available data may require converting to a standard unit, e.g. marginal metabolic equivalent of task (MMET) h/week (19) — the associated uncertainty might be quantified using a regression model fitted to data where both the reported unit and the standard unit were observed. Bayesian latent variable models have also been used in models of physical activity interventions (20) and for dietary exposures (21).

**Combining data sources**

If there are data from multiple sources on the same parameter, they may be combined using statistical methods such as meta-analysis. Multiple data sources could vary in how representative they are of the quantity of interest. For example, suppose we do not have data for our setting of interest, but we do have data for multiple similar settings, then the observed variation between settings indicates the uncertainty about the setting of interest. This principle, illustrated in our ITHIM example, can be expressed formally through a hierarchical model (17,22), also known as a multilevel, random effects or mixed effects model. A hierarchical meta-analysis is used in the example to estimate the background air pollution concentration for a city by combining a weak prior distribution with data from other cities in the region, leading to an assessment of uncertainty that draws on the available data. If predictors of differences between settings are recorded, then corresponding regression terms may be included in the model. In general, a hierarchical model can be designed to reflect beliefs about any similarities or biases that are expected in particular data sources, for example, meta-analyses can be bias-adjusted to downweight studies thought to have lower rigour or relevance (23).

Sometimes a quantity of interest will vary smoothly and continuously between settings, for example, changes in exposure to risk over time or space, or between people of different ages. Information from the point of interest may be strengthened by combining with nearby measurements, while observations that are too distant might simply be excluded. There is a large literature on spatial statistical modelling of disease exposures and outcomes, see. e.g. (24,25). A typical approach is to represent correlations between area-level outcomes through random effects as part of a hierarchical model. These effectively downweight data that are further away from the place of interest — our example might have been extended in this way. Similar models can represent correlations in time. Non-linear regression can be used to explicitly model how a quantity depends on time or age. This allows weak data from a particular time to be strengthened using data from nearby times, under the assumption that the time trend follows some functional form. Splines and related methods (26) can model trends both flexibly and parsimoniously, though we must be judicious in extrapolating beyond the data.

**Choosing a conservatively wide distribution in the absence of data**

Sometimes there will be no data to inform a model parameter, or we may be not be willing to make the assumptions required to use any indirectly related data. To enable VoI methods, we recommend that a conservatively wide distribution is chosen for the parameter which might overestimate, but is unlikely to underestimate, the extent of uncertainty (27,28). A low VoI would then allow further research on this parameter to be confidently ruled out. Further research for a weakly informed parameter might involve structured expert elicitation (see (29) for instruction, or (30) for an air-pollution–specific example).

# Value of Information

Value of Information (VoI) analysis measures the extent to which reduction of uncertainty in model parameters reduces uncertainty in the output of a model. It is used in health economic modelling (6), where the purpose of the model is to choose between a set of policy decisions $d = 1, \ldots, D$. The optimal decision minimises the expected loss $L(d, X)$ (or maximises the net benefit) under current information about $X$. The "current information" is quantified by the uncertainty distributions defined for $X$. The expected loss under the optimal decision is therefore $min_d E_X\{L(d, X)\}$, which is expected to reduce when further information is collected to reduce uncertainty around $X$.

Here, we use the *expected value of partial perfect information* (EVPPI) for a particular parameter or group of parameters $X_i$, which is the expected reduction in loss if we were to learn $X_i$, while the other parameters $X_{-i}$ remain uncertain:

$$EVPPI_{X_i} = \min_d E_X\{L(d,X)\} - E_{X_i}\left[\min_d E_{X_{-i}|X_i}\{L(d, X_i, X_{-i})\}\right]. \quad (1)$$

However, in health impact models, there is not always a decision between finite options. The aim is commonly to estimate the expected health gains $Y = f(X)$ from a potential scenario, under uncertainty. Decision-theoretic principles can still be used to obtain the EVPPI (9). The "decision" $d$ in this case is the choice of estimate $\hat{Y}$ for $Y$, and the "loss" can be taken as the inaccuracy of this estimate. Specifically, under a *squared error* loss $L\left(\hat{Y}, Y\right) = \left(\hat{Y} - Y\right)^2$, the optimal estimate is $\hat{Y} = E_X(Y)$, the mean of $Y$ with respect to uncertainty about $X$, which can be estimated by Monte Carlo, as the average of $R$ simulations $Y^{(r)} = f(X^{(r)}), r = 1, \ldots, R$, where $X^{(r)}$ are samples from the uncertainty distribution of $X$. The expected loss under current information is the *variance* of $Y$, that is $var_X(Y) = E_X$.

The EVPPI for $X_i$ is then the expected *reduction in variance* in $Y$ if $X_i$ were to be learnt precisely.

$$EVPPI_{X_i} = var_X(Y) - E_{X_i}\left(var(Y|X_i)\right) \quad (2)$$

This can be interpreted as the amount of uncertainty in $Y$ explained by uncertainty about $X_i$. This could also be expressed as a percentage of variance explained, $100 \times EVPPI_{X_i}/var_X(Y)$, or as a predicted reduction in standard deviation, as $\sqrt{var_X(Y)} - \sqrt{E_{X_i}(var(Y|X_i))}$.

This method has been used for sensitivity analysis in computer modelling (10), and, as we have described, it has a Bayesian decision-theoretic justification. The parameter with the highest EVPPI is interpreted as the one for which obtaining extra data would lead to the greatest improvement in precision for the estimate of health gains. In our ITHIM example, the output of interest is the expected number of premature deaths averted in the "best case" air pollution reduction scenario.

If there are $R$ outcomes of interest $Y_r: r = 1, \ldots, Y_R$, then we could calculate a different VoI for each $Y_r$. Alternatively we could define a "generalised variance" to use in place of the variance in the definition of EVPPI, that gives an overall measure of uncertainty about all outcomes simultaneously, e.g. as the sum of the variances, potentially weighted if some outcomes are judged to be more important than others. See (9) for more details.

The EVPPI expressions in Equations 1 and 2 above can be estimated, given a sample of values from the uncertainty distributions of the model inputs, and a corresponding sample for the model outputs, by using regression. This is illustrated in the Appendix — note that this can be done in two lines of R code.

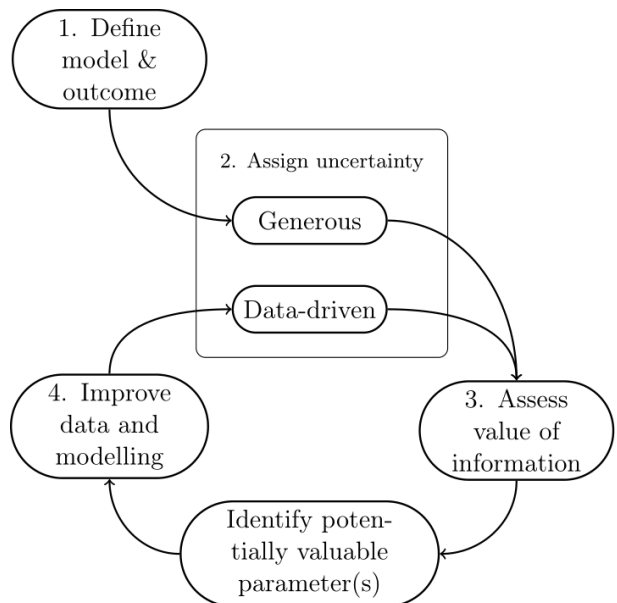

Figure 2. Model improvement and research prioritisation through Value of Information methods

# Using VoI to prioritise model improvements

Research to obtain additional data, and any subsequent statistical modelling , require time and resources, and may not ultimately affect the final results. VoI methods can be used to guide model development by targeting components of the model for which further data is expected to improve the precision of the final results. We recommend the following process, illustrated in Figure 2:

1. Develop the model structure, including every potentially important quantity as a parameter.

2. Define the outcome(s) of interest, or the decision, arising from the model.

3. Obtain distributions representing uncertainty about each parameter, based on available data, using a generous or conservatively wide distribution if the choice is unclear. If informal sensitivity analysis suggests a parameter value might not affect the results of the model, it can be excluded from VoI analysis.

4. Calculate EVPPI for each model parameter. If any have low EVPPI, so that better information on them would not strengthen the final result of the model, then further research on them is not worthwhile for this model.

5. Return to Step 3 and undertake further data collection and modelling to inform any parameters with substantial EVPPI.

The judgement of what EVPPI is deemed substantial enough to deserve a particular investment of research time could be made informally, as in our example. In health economic modelling, this could be done formally by quantifying the expected health benefits from reduced decision uncertainty in economic terms, and trading off with research costs (6). Additionally, if we are in a position to conduct a study to collect primary data, the *expected value of sample information* could be used to quantify the expected gains from a study of a specific design and sample size, though this is not illustrated here — see, e.g. (9) for an example.

**Health impact models with multiple model variants**

We emphasise that the value of information depends on the chosen *model structure* (Step 1) and *outcome* of interest (Step 2). This is illustrated in Figure 3. In our example, we examine a model that quantifies health impacts of air pollution scenarios in São Paulo. However there are multiple variants of that model designed to assess impacts for other cities and countries. Each setting requires slightly different data. Some quantities, such as patterns of exposure to pollution, may vary between places, whereas other quantities, such as the relative risks of disease due to exposure, are assumed to be generic to all settings. Collecting further data on such generic quantities would benefit all variants of the model, therefore estimating the *full* value of such data would require an analysis based on an expanded model structure that includes all settings.

Likewise there are typically multiple health outcomes of interest (e.g. deaths averted, reductions in health inequality, reduction in burden of a particular disease) and many alternative scenarios to examine. If the model is used for an explicit *decision* between policies, then we can define VoI as in Equation 1. If the model is just used for *estimation*, we could calculate a different VoI (Equation 2) corresponding to each outcome and scenario, or define the VoI based on some overall measure of uncertainty.

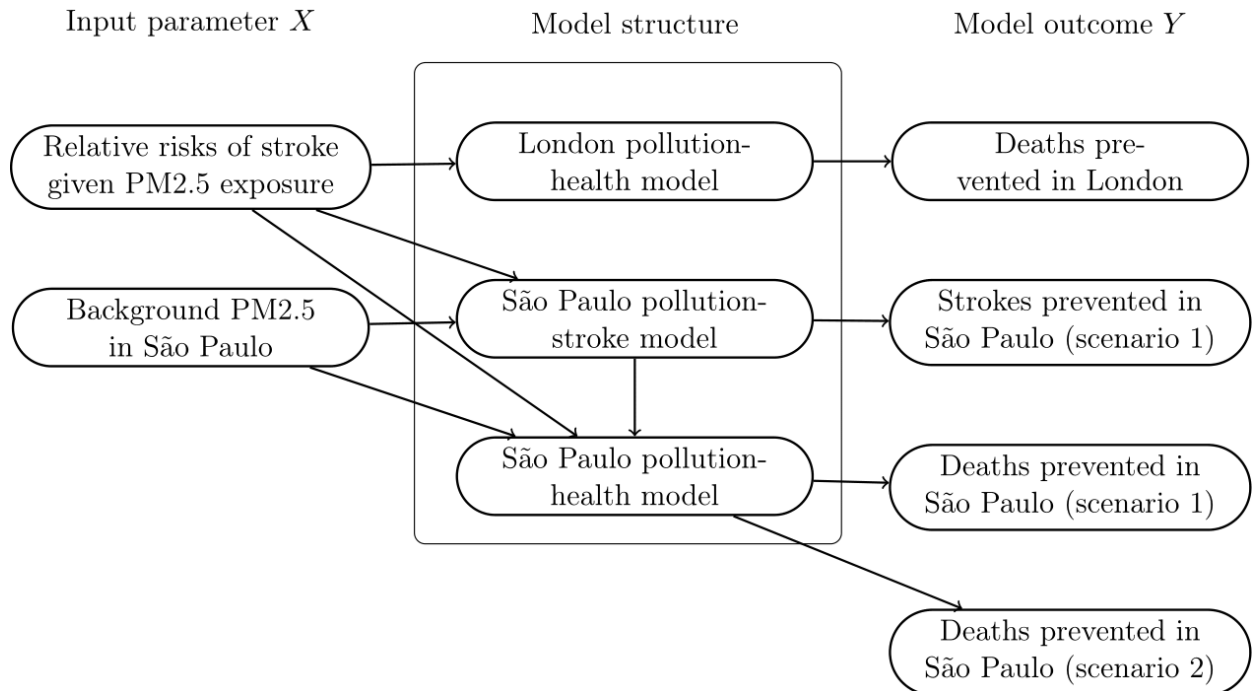

*Figure 3. Illustration of how the value of learning a particular input parameter depends on the choice of model structure and outcome of interest. In this hypothetical model, learning the relative risk of stroke associated with pollution improves information about health gains in São Paulo and London, whereas learning the air pollution concentration in São Paulo will only improve information on São Paulo-specific outcomes. The expected value of information also depends on the definition of the disease outcome of interest, and the air pollution reduction scenario examined.*

## Using VoI to guide improvement of a health impact model

Here we demonstrate how VoI can be used to prioritise model improvements, in a hypothetical example, which is fully described in the Appendix. The example is adapted from the published ITHIM São Paulo (14) by using the original model structure, but starting with very broad distributions on the parameters (explained in the Appendix), in order to imitate the typical development of a health impact model in a new setting where an appropriate structure has been decided but the parameters are not known. We then show how VoI methods could be used to prioritise which parameters should be learnt better, and how hierarchical models can be used to inform parameters from multiple sources of indirect data. The example here is artificial and serves to demonstrate the methods, rather than as a substantive and thorough health impact analysis.

The model estimates the change in the burden of disease due to pollution exposure, for each disease and scenario, by age group and gender. To apply VoI, we choose the outcome of interest to be the reduction in the number of deaths per year from any cause, over all age and gender groups, under a scenario ("SP2040") defined as the "best case" for São Paulo in 2040 (14). We follow the process above, calculating the EVPPI for each parameter as the expected reduction in variance of this outcome if the parameter were to be learnt precisely. The EVPPI for each parameter is compared to the expected costs of research (in person–days) to obtain better information on the parameter. Research is undertaken if the potential precision reduction was thought to be worth the research cost, according to an informal judgement.

For all the parameters in our example, further research is judged to be possible that would involve searching literature for additional data and constructing a statistical model (Table 1). This kind of research is predicted to take 2–5 person-days per parameter in our example, but in other examples this may vary. In other situations, research of this kind may not be possible, and obtaining further data would require primary data collection (e.g. local air-pollution monitoring), which would be substantially more expensive.

For our example, we informally allocate a research "budget" of ten person–days. With this budget, we aim to maximise expected reduction in outcome uncertainty. The costs of research, and the actual amount of information (compared to "perfect information") gained from the research will not be possible to predict exactly in practice; therefore, informal judgements are usually required to make this trade-off. Although the need to make informal judgements may make the approach sound less scientific, it is important to remember that modelling requires laying out clearly assumptions made regarding data and evidence. The alternative would be to make such decisions implicitly without laying out the assumptions (31).

# Results

Under the initial set-up of the model, we estimate the expected number of deaths prevented per year in the SP2040 scenario to be 189.4, with a standard deviation of 196.0.

The EVPPI and the expected cost of further research for each parameter are presented in Table 1. The two parameters with the highest EVPPI are $\zeta$, the proportion of air pollution attributable to road traffic, and $\eta$, the background level of air pollution, both pertaining to São Paulo. As we anticipated that five days of work would provide substantially better information for each parameter, and given our ten person–day budget, we choose to obtain further data on these parameters.

Learning the exact value of $\zeta$ could be expected to reduce the variance of the number of deaths averted by up to 48%, from $38,000 \approx 196.0^2$ to $20,000 \approx 140.7^2$, while learning $\eta$ would be expected to reduce the variance by 12%. Literature is then searched for data, followed by statistical modelling, to obtain better-informed estimates of $\zeta$ and $\eta$ respectively. The data to inform $\zeta$ and $\eta$ consist of alternative published estimates of the proportion of air pollution from road traffic within São Paulo, and the data for $\eta$ consist of estimates of background pollution for other cities and countries. Bayesian hierarchical modelling is used to produce posterior distributions quantifying the updated beliefs about $\zeta$ and $\eta$, informed by the variability between the published estimates. Full details of these models are given in the Appendix.

After this research, we estimated the number of deaths prevented to be 223.3, with standard deviation 132.4, which is judged to be sufficiently precise, given the likely cost of further research.

Table 1. Expected value of partial perfect information (expressed in two different ways) and expected amount of research required for improved information, for each parameter in the example air pollution health impact model .

| Parameter | Description | EVPPI expressed as: | | Expected person–days of work for improved information |
|---|---|---|---|---|
| | | Predicted reduction in standard deviation describing uncertainty in number of deaths prevented | Percentage of outcome variance explained by parameter uncertainty | |
| $\eta$ | Background air pollution | 12.2 | 14.68 | 5 |
| $\zeta$ | Traffic proportion of air pollution | 55.2 | 48.20 | 5 |
| $\alpha$ | Proportions of traffic pollution emissions from each mode | 3.4 | 7.53 | 5 |
| $\lambda_{walk}$ | Walking ventilation rate | 9.4 | 7.50 | 3 |
| $\lambda_{cycle}$ | Cycling ventilation rate | 0.3 | 0.43 | 2 |
| | Uncertainty in dose–response curve for relative risk of disease as a function of air pollution: | | | |
| $\xi_{stroke}$ | Stroke | 0.3 | 3.82 | 5 |
| $\xi_{IHD}$ | Ischaemic heart disease | 2.7 | 3.35 | 5 |
| $\xi_{LC}$ | Lung cancer | 0.2 | 1.38 | 5 |
| $\xi_{COPD}$ | COPD | 1.7 | 0.68 | 5 |

# Discussion

We have presented a procedure for prioritising research to reduce uncertainty in health impact simulation models. VoI methods are used to identify which model inputs contribute most to the uncertainty about the outputs. This allows the model developer to decide where to focus resources to improve the model, given the cost of doing research for each model input. Such methods might be particularly helpful for prioritising research to inform policy making in contexts with limited data (e.g. low and middle-income countries). In the pollution and health example, VoI identified two particularly uncertain parameters describing pollution levels and sources, which were subsequently informed using hierarchical modelling of related data, leading to estimates of health impacts with a 50% improvement in precision.

Once uncertainties have been quantified probabilistically, no further assumptions are necessary to obtain the potential value of further information on each parameter, which follows naturally through decision-theoretic principles. A more common approach to research prioritisation is one-way sensitivity analysis, that is, comparing a model output for alternative values of the input. This depends on an arbitrary choice of alternative values, and may be misleading if the relationship of the input to the output is nonlinear. While simple sensitivity analysis is necessary if we are not able to even roughly quantify the true extent of uncertainty about a model input, we would argue that statistical methods, such as hierarchical models, can be used to quantify uncertainty in many more cases than in current practice.

Note that the estimates of VoI are conditional both on the model structure, outputs of interest, and on the distributions chosen for the input parameters. If the model structure is uncertain, it could be extended to include parameters that represent different plausible structural assumptions, as in (27). This allows VoI to be used to prioritise research to refine the model structure. In our case study, VoI analysis was based on a single output of interest. More experience is needed on using VoI for research prioritisation when there are several outputs of interest, and several modelling settings and scenarios to be considered.

The models we have described estimate population-average outcomes as a complex function of a set of input parameters. VoI analysis is enabled by defining distributions that quantify uncertainty on the input parameters. As outlined in (32), there are many alternative structures for health policy simulation models, and one of these is "microsimulation", which works by simulating outcomes for a large population of synthetic individuals, and then taking an average. In these, uncertainty is sometimes not expressed explicitly by placing distributions on inputs, but implicitly, through using a finite sample of observed individual-level data to generate the synthetic population (see (33) for an example). Calculating the value of information in this context would be challenging — though could be done in theory by defining a parametric model from which synthetic individuals could be sampled.

Note that the consequences of overestimating and underestimating uncertainty are different. If uncertainty for an input is overestimated, but the VoI is still low, we can be confident that further research on that input would not be worthwhile, whereas if the VoI is high, we will obtain better evidence. On the other hand, if we underestimate uncertainty, VoI analysis might tell us wrongly that further research is not needed. Thus we recommend being 'generous' or 'conservative' in assigning uncertainty to parameters, if the priority is to obtain a better evidence base for decision making. Any new evidence obtained from research may have value beyond the decision model it was originally obtained for.

# A guide to Value of Information methods for prioritising research in health impact modelling: Appendix.

## 1. The pollution and health impact model

Code to implement this example is available at https://github.com/robj411/value_of_information_example.

The calculations relating the model input parameters to the expected number of deaths averted in the example pollution and health impact model, adapted from (14) are given in this appendix. Greek letters denote parameters that are uncertain. Upper-case Roman characters refer to data sets, computed quantities, and functions. Many of these quantities are computed by subgroups, defined by combinations of five variables, indexed by lower-case Roman characters as described in Table 1.

*Table 1. Indices used in the ITHIM variables*

| Label | Values |
|---|---|
| $a$ | Age group |
| $d$ | Disease |
| $g$ | Gender |
| $m$ | Transport mode: |
|  | Walking |
|  | Cycling (cyc.) |
|  | Bus |
|  | Car |
|  | Motorbike (mot.) |
|  | Goods vehicles (GV) |
| $s$ | Scenario |

## Parameter values

The model structure from (14) was used, but we start afresh with choosing parameter values, in order to emulate a typical situation in health impact modelling where we wish to use an existing model structure in a new setting, and to demonstrate how Value of Information analysis might then be used to prioritise research to inform the model parameters.

Initially, we choose prior distributions for parameter values that cover the range we believe each value could plausibly take. For example, for $\eta$ and $\zeta$, we can use World Health

Organisation (WHO) tables that include data from all over the world to guide us towards what values would be plausible.

The background PM2.5 concentration for a city, $\eta$, ranges between 1.6–217, with all cities in Latin American countries having values below 50, so we use a log-normal distribution with mean 3 and standard deviation 1 on the log scale (Figure 1, right, grey). The proportion of PM2.5 pollution attributed to transport in a related WHO database ranges from 0% to 85% (including dust, the range is 0% to 91%). The value given for São Paulo is 23.5% (51%, including dust). Thus, we choose Beta(2,3) as the prior distribution for $\zeta$ (Figure 1, left, grey).

For $\alpha$, we first consult a recent CETESB report[1] for the Metropolitan area of São Paulo. These are estimates of contributions to particulate matter of different vehicle types. Cars, buses, motorcycles and goods vehicles are estimated to contribute 32%, 4%, 4%, and 60%, respectively. A Dirichlet(32, 4, 4, 60) distribution is used as the initial belief about the vector $\alpha$ of these four proportions, which encodes a prior expectation centred around these published values, but with some uncertainty, as illustrated in Figure 2.

For the $\lambda$ values, we consult The Compendium of Physical Activity[2]. Estimates for walking METs vary from 2 to 12, and for cycling from 3 to 17. For our model, we use marginal METs (MMET = MET – 1). For these values we choose log-normal distributions, so that they are greater than zero and peak earlier in the range. These distributions are shown in Figure 3.

The parameters $\xi_d$ express uncertainty about the dose–response curves $H_d(x)$ (Figure 4), defined as the relative risk of mortality from each of four diseases $d$ associated with an increase of $x$ in air pollution exposure. They are assumed to be measured with some degree of uncertainty, quantified by $\xi_d$. Specifically the true relative risk is defined as $\xi_d H_d(x)$, where $\xi_d$ has a log-normal distribution with mean 0 and standard deviation 0.5, thus it is centred around 1. This represents the belief that the relative risk does not go below 1, and that the actual value is between 50% and 200% of the reported values (see Figure 4).

The average time spent travelling, the population composition, and the baseline burden of disease are treated as constants for the purpose of this illustration, though in practical applications they might be treated as uncertain.

---

[1]     Table 5 of https://cetesb.sp.gov.br/veicular/wp-content/uploads/sites/6/2013/12/Plano_de_Controle_de_Poluicao_Veicular_do_Estado_de_Sao_Paulo_2014-2016.pdf, where we include "business" in the "heavy goods" category.

[2]     https://sites.google.com/site/compendiumofphysicalactivities/references

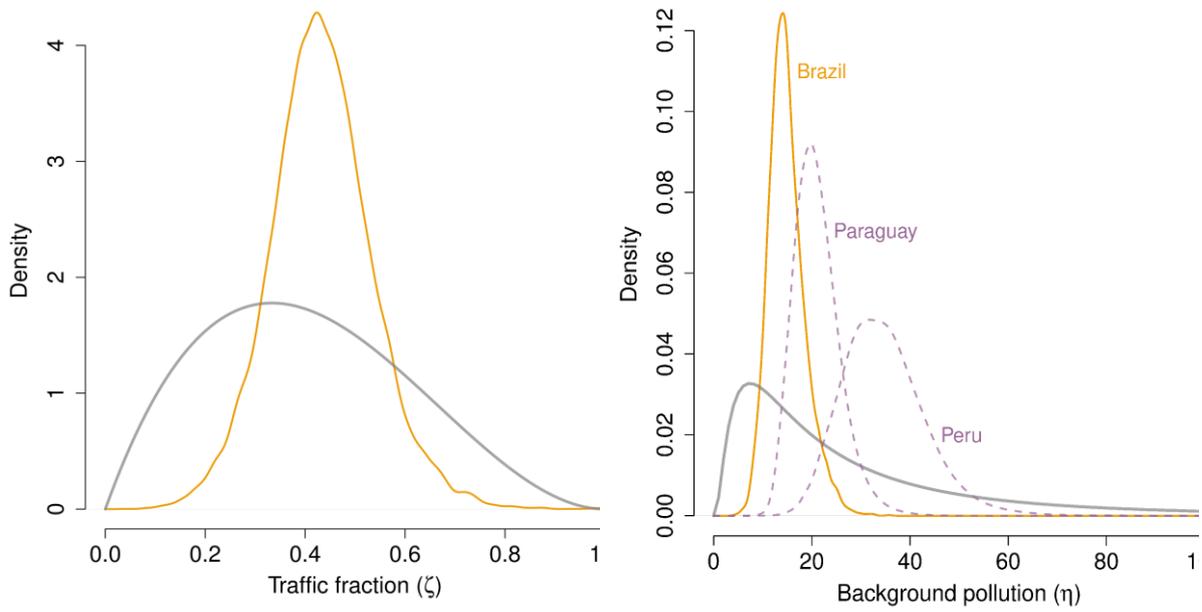

*Figure 1. Distributions for ζ and η in the pollution module. In grey are the original vague prior distributions: left, the proportion of air pollution attributed to traffic, ζ ~ Beta(2,3), and right, the mean background air pollution level for cities within Brazil, η ~ Lognormal(3,1). In orange are the posterior distributions given the additional data. The right panel also shows the posterior distributions for background pollution for predicted cities in two other countries in the lower–middle income American region: Peru and Paraguay.*

*Table 2. Parameters of the pollution and health impact model, with prior distributions*

| Name | Description | Prior distribution |
|---|---|---|
| | | |
| $T_{a,g,m,s}$ | Minutes spent travelling per day | Constant |
| $N_{a,g}$ | Population number | Constant |
| $U_{a,d,g,o=death}$ | Background burden of disease: deaths per age, gender and disease | Constant |
| | | |
| $\eta$ | Background PM2.5 concentration | Log-normal(3,1) |
| $\zeta$ | Fraction of $\eta$ attributed to road transport | Beta(2,3) |
| $\alpha$ | Fraction of $\zeta$ attributed to four different modes (bus, car, motorcycle, goods vehicles) | Dirichlet(32,4,4,60) |
| | | |
| $\lambda_{m=walk}$ | Walking MMET[3] | Log-normal(1,1) |
| $\lambda_{m=cyc.}$ | Cycling MMET | Log-normal(2,0.4) |
| | | |
| $H_d(x)$ | Relative risk of disease (Figure 1) | Function of PM2.5 exposure |
| $\xi_{d=ALRI}$ | Uncertainty in ALRI[4] relative risk | Constant (1) |
| $\xi_{d=COPD}$ | Uncertainty in COPD[5] relative risk | Log-normal(0,0.5) |
| $\xi_{d=LC}$ | Uncertainty in lung cancer relative risk | Log-normal(0,0.5) |
| $\xi_{d=stroke}$ | Uncertainty in stroke relative risk | Log-normal(0,0.5) |
| $\xi_{d=IHD}$ | Uncertainty in IHD[6] relative risk | Log-normal(0,0.5) |

---

[3] Metabolic equivalent, kcal/kg/hr (41)

[4] Acute lower respiratory illness

[5] Chronic obstructive pulmonary disease

[6] Ischaemic heart disease

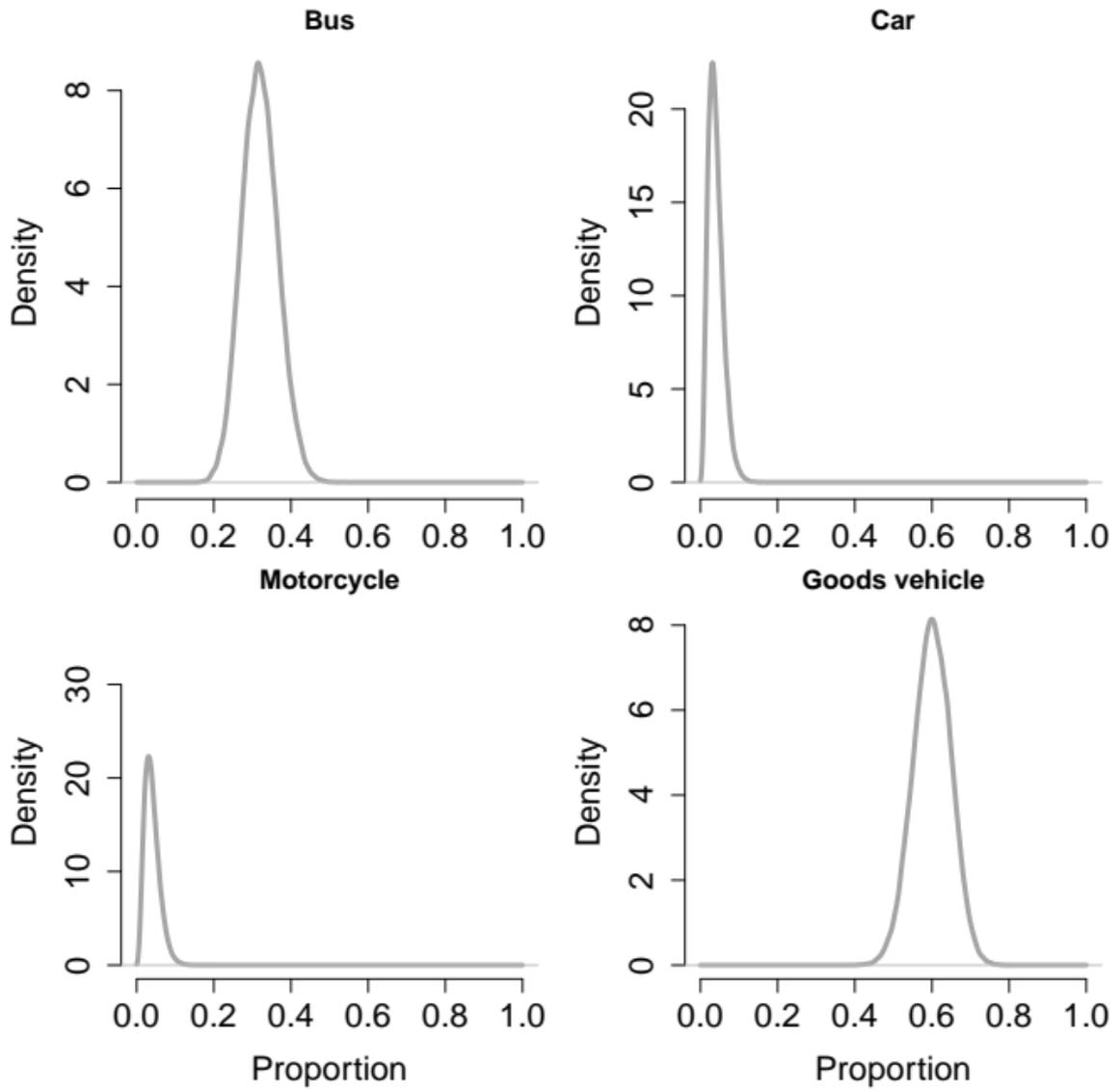

*Figure 2. Distributions for α, the proportion of traffic pollution attributed to each mode.*

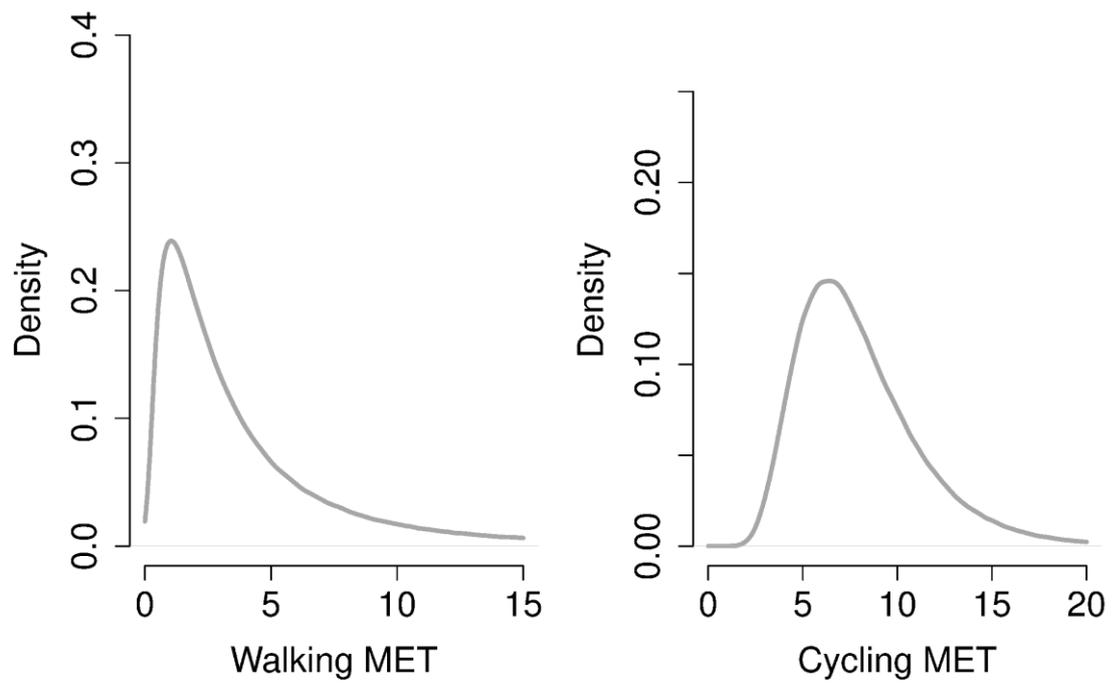

*Figure 3. Distributions for λ, the MMET values associated with active travel.*

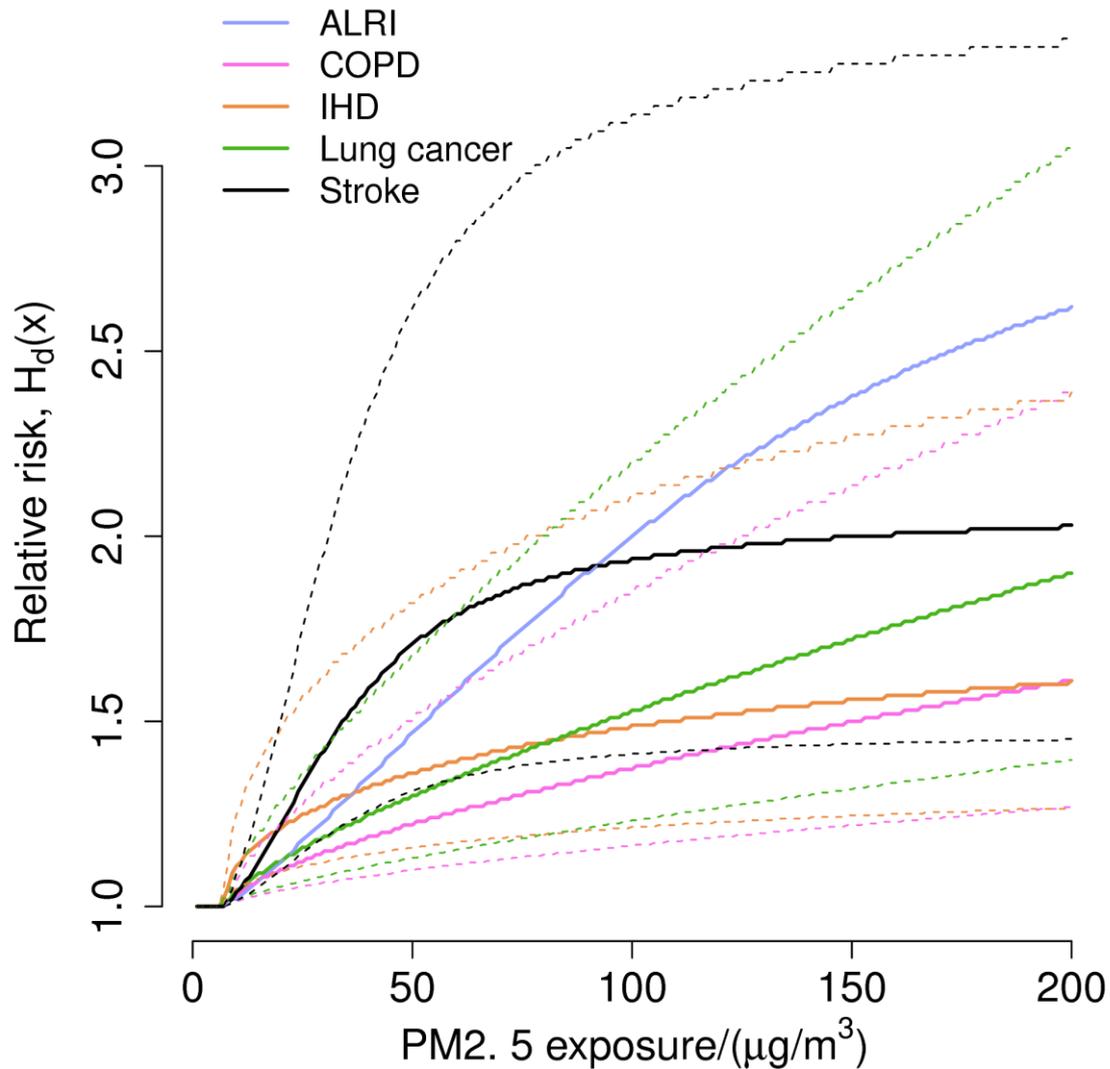

*Figure 4. Relative risk of disease, $H_d(x)$, for five diseases $d$ and PM2.5 exposure $x$ from zero to 200, from (34). $H_d(x)$ expresses the risk of disease $d$ between two individuals whose PM2.5 exposure differs by $x$. The value for $x$ will be different in each scenario; the quantity necessary to calculate the burden of disease is the value of $H_d(x)$ relative to the baseline scenario value. This relative risk multiplies the background burden of disease to give the scenario burden of disease. Dashed lines show 90% uncertainty intervals implied by the parameters $\xi_d$.*

### Model calculations

Here we describe all calculations in the model that map the input parameter values to the outcome, which is the disease burden in terms of the number of deaths.

### Population travel

First, the total time spent travelling is a product of the population numbers and the average time spent travelling per demographic group, per mode of transport and scenario.

$$A_{a,g,m,s} = N_{a,g} \cdot T_{a,g,m,s}.$$

Then the total time spent travelling per mode and scenario is the sum over the demographic groups:

$$\tilde{A}_{m,s} = \sum_{a,g} A_{a,g,m,s}.$$

Finally, we calculate the total time spent travelling per mode and scenario relative to the baseline:

$$\hat{A}_{m,s} = \tilde{A}_{m,s}/\tilde{A}_{m,s=baseline}.$$

*Air pollution calculations*

First, we calculate the fold change in air pollution by mode, given the population travel pattern, for the motorised modes which contribute to pollution emissions, $m \in \{bus, car, mot., GV\}$:

$$\tilde{P}_{m,s} = \hat{A}_{m,s} \cdot \alpha_m,$$

and sum over the modes to get the fold change in total transport-related air pollution:

$$\hat{P}_s = \sum_m \tilde{P}_{m,s}.$$

We use this to calculate the fold change in background concentration of air pollution, given the proportion attributable to traffic:

$$\overline{P}_s = \eta \left(1 - \zeta \left(1 - \hat{P}_s\right)\right).$$

*Ventilation rate calculations*

We define ventilation rates for the different modes, relative to ventilation when not travelling, as follows:

$$\begin{array}{ll} V_{m=walk} & 1 + \lambda_{m=walk}; \\ V_{m=cyc.} & 1 + \lambda_{m=cyc.}; \\ V_{m=bus} & 1.5; \\ V_{m=car} & 1.5; \\ V_{m=mot.} & 2; \\ V_{m=GV} & 1.5. \end{array}$$

Then, for each mode, we calculate the total ventilation rate in traffic, given each demographic group's travel pattern:

$$\tilde{V}_{a,g,m,s} = V_m \cdot T_{a,g,m,s}.$$

The total ventilation rate per demographic group per scenario sums over time spent in traffic and that spent not in traffic:

$$\hat{V}_{a,g,s} = \left(1{,}440 - \sum_m T_{a,g,m,s} + \sum_m \tilde{V}_{a,g,m,s}\right)/1{,}440.$$

Finally, the exposure to air pollution is calculated as the product of ventilation and background air pollution concentration:

$$\check{V}_{a,g,s} = \hat{V}_{a,g,s} \cdot \overline{P}_s.$$

## Health impact calculations

Health impacts are calculated by standard comparative risk assessment methods (see, e.g. (35)). We calculate the relative risk of disease given air pollution exposure as

$$\hat{H}_{a,d,g,s} = 1 + \xi_d \cdot \left(H_d\left(\check{V}_{a,g,s}\right) - 1\right),$$

where $\xi$ defines the uncertainty about the dose–response relationship.

We calculate the relative risks between scenario $s$ and baseline as

$$\check{H}_{a,d,g,s} = \hat{H}_{a,d,g,s} / \hat{H}_{a,d,g,s=baseline}.$$

This ratio is used to scale the baseline burden of disease, to obtain the disease burden for each scenario, as follows:

$$\tilde{U}_{a,d,g,s} = \check{H}_{a,d,g,s} \cdot U_{a,d,g}.$$

Finally, we sum over the demographic groups to get the total burden of disease, which we count in number of deaths from the avoidable causes that are modelled, for the population of Sao Paulo, per year.

$$death_s = \hat{U}_s = \sum_{a,d,g} \tilde{U}_{a,d,g,s}.$$

## 2. Calculating the expected value of partial perfect information (EVPPI)

The EVPPI can be estimated, given a sample of values from the uncertainty distributions of the model inputs, and a corresponding sample for the model outputs, by using regression. The method of (13) is used here, as adapted in (9) for cases where the EVPPI is defined as an expected reduction in variance. The following regression model is fitted, where $Y^{(r)}$ are the sampled values of the outcome $Y$, $X_i^{(r)}$ are the corresponding samples of the input parameter(s) of interest $X_i$, $\epsilon_r$ is a noise term, and $g(\ )$ is a flexible, non-linear function.

$$Y^{(r)} = g\left(X_i^{(r)}\right) + \epsilon_r, r = 1, \ldots, R$$

Then $E_{X_i}(var(Y|X_i))$ is estimated by the mean of the squared residuals from this regression, where $\hat{g}(\ )$ is the fitted regression function. For $g(\ )$, we use generalised additive models, via the `gam` function in the `mgcv` R package, as in the following R code. $Y$ is the vector of length $R$ containing the sample of model outputs, and $Xi$ is the corresponding vector of sampled values for the input parameter of interest.

```
library(mgcv)
g.hat <- gam(Y ~ s(Xi))
EVPPI.xi <- var(Y) - mean((Y - g.hat$fitted)^2)
```

To learn the expected value of learning more than one parameter simultaneously, for example two parameters $Xi,Xj$, the univariable regression is extended to a multivariable regression:

```
g.hat <- gam(Y ~ te(Xi,Xj))
```

Note that the method in (36) calculates this expected variance reduction via a different route, though is not easily applicable (9) when Markov Chain Monte Carlo simulation is used to estimate the parameters of a Bayesian statistical model, as in our example.

## 3. Meta-analysis of proportion of air pollution attributable to road traffic

We obtain information about $\zeta$ using three data sources, giving estimates of: 40% (37), 30% (38), and 60% (39), labelled $y_i$ for studies $i = 1,2,3$. The data were produced in different years (2007–8, 2003 and 2011–2014 respectively), at three different locations using three different methods. The quantity of interest for our model is a city-level average for the year 2016.

We construct a Bayesian hierarchical meta-analysis model (40) to relate this quantity to the observed data. For simplicity, the observed estimates $y_i$ are assumed to be logit-normally distributed around the true city-level average $\zeta$, that is $logit(y_i) \sim N(logit(\zeta), \tau^2)$. We use our originally assigned distribution for the prior distribution, which is $eta(2,3)$, representing a belief that $\zeta$ is between 6% and 81% with 95% probability (Figure 1). The between-source standard deviation on the logit scale, $\tau$, is given a positive-truncated Normal(0,1) prior, which represents a belief that $\tau$ is between 1% and 30% with 95% probability.

Samples from the resulting posterior distribution of $\zeta$ are generated by Markov Chain Monte Carlo methods, using the software Stan (42). The prior and posterior densities are illustrated in Figure 1 (left). Given only three observations, the posterior will be influenced by the exact choice of prior. However Figure 1 (left) shows that it acknowledges that the true $\zeta$ may be similar to any of these three observations. This rough analysis is sufficient for research prioritisation — a more detailed analysis might account for observed characteristics of each data source, e.g. using regression, and explore alternative statistical assumptions.

# 4. Hierarchical models to estimate the background air pollution concentration

We suppose we do not have any direct data for the background concentration of PM2.5, $\eta$, for São Paulo. Instead, we use World Health Organisation (WHO)--published estimates for the background PM2.5 concentrations $\eta$ g/$m^3$. We use data from 2016 that cover 2,972 cities worldwide.[7] We illustrate how a hierarchical model (also known as a multi-level or mixed effects model) can make use of this information to form an uncertainty distribution for $\eta$.

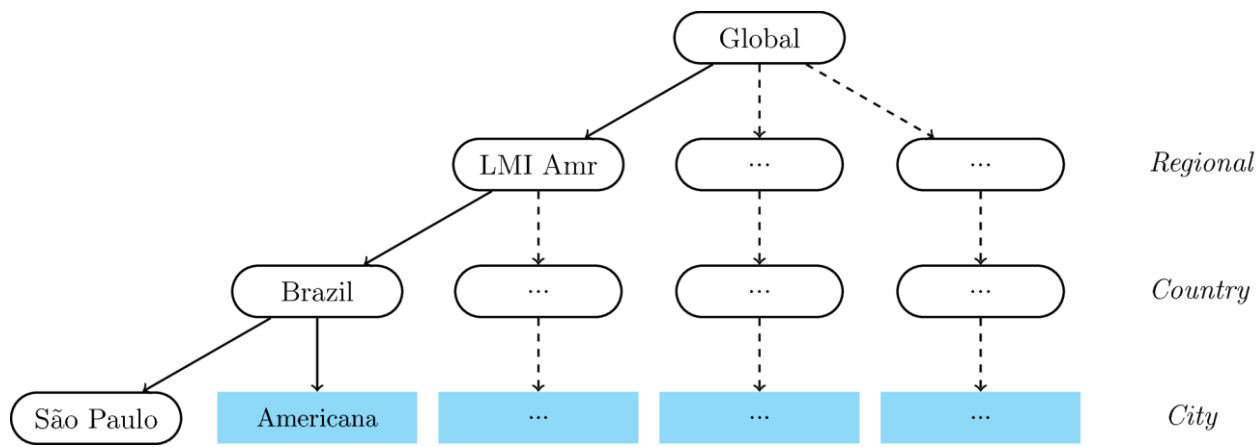

*Figure 5. The hierarchical model used to infer the background level of air pollution for São Paulo. Blue boxes represent measurements from cities all over the world, which are related by the country they are in, which are related by the region they are in. Ovals represent distributions that we will learn from the data. Ellipses denote additional locations not shown. There are, in the dataset, ten regions, 104 countries, and 2,972 cities.*

Each city $i$ belongs to one of 104 countries $c$, and each country belongs to one of ten regions $r$. Brazil is in the lower–middle income Americas region (LMI Amr), which contains 102 measurements from cities in 13 countries (see *Figure 5*). The published city-specific estimates $y_{icr}$ are assumed to be generated from a (log-)normal distribution centred around country-level means $\theta_{cr}$. A second and third level of the model similarly define country-level means in terms of a regional mean $\theta_r^{(R)}$, and the regional means in terms of a global mean $\theta^{(G)}$. Formally:

---

[7] http://www.who.int/phe/health_topics/outdoorair/databases/cities/en/

$$\begin{aligned}
\theta^{(G)} &\sim N(0, 5^2) \\
\theta_r^{(R)} &\sim N\left(\theta^{(G)}, \left(\tau^{(G)}\right)^2\right), r = 1, .., N^{(R)} \\
\theta_{cr}^{(C)} &\sim N\left(\theta_r^{(R)}, \left(\tau^{(R)}\right)^2\right), c = 1, .., N_r^{(C)} \\
\log(y_{icr}) &\sim N\left(\theta_{cr}^{(C)}, \left(\tau^{(C)}\right)^2\right), i = 1, .., N_{cr} \\
\log(\eta) &\sim N\left(\theta_{cr}^{(C)}, \left(\tau^{(C)}\right)^2\right) for\, c, r\, indicating\, Brazil
\end{aligned}$$

Standard half-normal prior distributions are used for the standard deviation parameters $\tau^{(C)}, \tau^{(R)}, \tau^{(G)}$, representing a 95% prior credible interval of about (0.03, 2.2) for the standard deviation at each level.

The unknown São Paulo value $\eta$ is assumed to be generated from the Brazil-specific country-level distribution. The posterior distribution of $\eta$ will principally reflect the variation between other cities in Brazil, but also include, less directly, information from the other regions and countries in the world. This is illustrated in Figure 1 alongside the posterior distributions for predicted cities in two other countries in the lower–middle income American region, Peru and Paraguay.

This relatively simple example serves to illustrate the basic principles — a more elaborate model might include information about observed characteristics of São Paulo that are predictive of the air pollution levels (e.g. size or population density), multiple measures from the same city, or correlations in space and time. See e.g. (43) for a sophisticated example of hierarchical modelling of global pollution.